\documentclass[pre,preprint,showpacs]{revtex4}

\begin{document}

\title{Scaling Analysis and Systematic Extraction of Macroscopic Structures
in Fluctuating Systems of Arbitrary Dimensions}
\author{Ning-Ning Pang$^*$}
\author{Hisen-Ching Kao}
\affiliation{Department of Physics, National Taiwan University, Taipei, Taiwan,
Republic of China}
\author{Wen-Jer Tzeng}
\thanks{Correspondence should be addressed either to NNP at
nnp@phys.ntu.edu.tw or WJT at wjtzeng@mail.tku.edu.tw.}
\affiliation{Department of Physics, Tamkang University, Tamsui, Taipei, Taiwan,
Republic of China}
\date{\today}

\begin{abstract}
Many fluctuating systems consist of macroscopic structures in addition to noisy signals.
Thus, for this class of fluctuating systems, the scaling behaviors are very complicated.
Such phenomena are quite commonly observed in Nature,
ranging from physics, chemistry, geophysics, even to molecular biology and physiology.
In this paper, we take an extensive analytical study
on the ``generalized detrended fluctuation analysis'' method.
For continuous fluctuating systems in arbitrary dimensions,
we not only derive the explicit and exact expression of macroscopic structures,
but also obtain the exact relations between the detrended variance functions
and the correlation function.
Besides, we undertake a general scaling analysis,
applicable for this class of fluctuating systems in any dimensions.
Finally, as an application,
we discuss some important examples in interfacial superroughening phenomena.
\end{abstract}

\maketitle

\section{Introduction}

Fluctuating systems have been a study of great interest
for their generic scaling behaviors widespread in Nature\cite{Krug97,Halpin95}.
Much attention has recently focused on interfacial superroughening phenomena
for their anomalous scaling behaviors,
different from the conventional scaling ansatz\cite{Sarma96}.
Many experiments have been performed, including growth of nickel surfaces
by pulse-current electrodeposition\cite{Saitou02},
cultivated brain tumor growth in a Petri dish\cite{Bru98},
spontaneous imbibition of a viscous fluid by a model porous medium\cite{Soriano05},
and molecular beam epitaxial growth\cite{Yang96,Jeffries96}.
All these superrough interfacial growth processes
consist of local interfacial orientational instability.
The boundary conditions (due to the finite size of real experiments)
restrict the infinite development of orientational instability.
Thus, the interfacial configuration gradually develops global mountains or valleys
as time increases.
The appearance of macroscopic structures is the signature
of superrough interfacial growth processes.
Interestingly, in a different context of science (computational molecular biology),
with some specific mapping rules the DNA sequences can be viewed
as a random walk\cite{Peng94,Buldyrev95}.
Lots of data show that the DNA sequences form mosaic structures (patchiness).
Peng et al.\cite{Peng94,Buldyrev95,Hu01} have proposed
a ``generalized detrended fluctuation analysis'' method to determine
the long-range correlation in DNA sequences
but exclude the effect caused by the mosaic structure.
The basic idea of this method is to eliminate the trend (or the macroscopic structure),
which is expressed in terms of a polynomial with the coefficients
determined by numerical least-squares-fit.
It has been adopted as a standard analysis method in various time series problems,
for example, the heart-rate fluctuations during sleep\cite{Bunde00},
interfiring time intervals between neural action potentials\cite{Bahar01},
and temporal fluctuations in seismic sequences\cite{Telsca01}.

However, few analytical works have been done
regarding the ``generalized detrended fluctuation analysis'' method.
Since many fluctuating systems are modeled by stochastic partial differential equations,
the explicit expression of the macroscopic structures (or the trends) is necessary
for further analytical study of such systems.
The numerical fitting can only give the values of various coefficients
but not their relations with the statistical quantities of the systems.
In addition, when a high-order detrending is necessary, the numerical least-squares-fit
of many coefficients consumes large amounts of computation time.
Besides, what are the explicit relations between the statistical quantities
of the systems before detrending and after detrending?
Can a general scaling analysis be derived?
Thus, we are strongly motivated to take an analytical study
on the fluctuating systems with formation of macroscopic structures.
In Ref. \cite{Pang04}, we have studied the fluctuating systems in one dimension.
In this paper, we plan to work out all these issues for fluctuating systems
in arbitrary dimensions.
Moreover, we notice that many superrough interfacial growth processes
consist of a very distinct property: the local roughness exponent equal to 1,
regardless of other exponents or parameters of the systems.
What is its origin?
What does it imply?
Are all these problems related?
In this paper, we also plan to explore this intriguing phenomenon in some detail.

The outline of this paper is as follows.
In Sec.\ II, we present the core ideas of
the ``generalized detrended fluctuation analysis'' method.
Then, we derive the exact expression of macroscopic structures in fluctuating systems
of arbitrary dimensions, expressed in terms of the Legendre polynomials.
In Sec.\ III, the exact relations between the correlation function
and the detrended variance functions are derived.
In Sec.\ IV, through the detailed investigation of the kernel functions,
we give a general scaling analysis applicable for any fluctuating systems
with macroscopic structure formation.
In Sec.\ V, as an application, we focus on some important examples
in interfacial superroughening processes.
In the last section, a brief summary is given.

\section{Systematic Extraction of Macroscopic Structures}

In this section, we will explicitly derive the exact expression
of the macroscopic structure of fluctuating systems in arbitrary dimensions.
First, let $y({\bf x})$ denote a random function of ${\bf x}$,
with ${\bf x}$ being a vector in the $d$-dimensional space.
One of the most important statistical quantities is the variance of $y({\bf x})$
in an observation window:
\begin{equation}
w^2\left(\prod_{i=1}^dl_i;\tilde{\bf x}\right) \equiv
\left\langle\left[y({\bf x})-\langle y({\bf x})\rangle_{\prod_{i=1}^dl_i;\tilde{\bf x}}
\right]^2\right\rangle_{\prod_{i=1}^dl_i;\tilde{\bf x}}
\end{equation}
with $\langle\ldots\rangle_{\prod_{i=1}^dl_i;\tilde{\bf x}}$
denoting the average over an observation window,
centered at $\tilde{\bf x}$ and of size $\prod_{i=1}^dl_i$.
The corresponding statistical quantity, related to macroscopic structure extraction,
is the $\{q\}$-th ($\{q\}\equiv\{q_1,\ldots,q_d\}$) degree detrended variance
of $y({\bf x})$ in that observation window:
\begin{equation}
w^2_{\{q\}}\left(\prod_{i=1}^dl_i;\tilde{\bf x}\right)\equiv
\left\langle\left[y({\bf x})-y_{\{q\}}({\bf x};\tilde{\bf x})\right]^2
\right\rangle_{\prod_{i=1}^dl_i;\tilde{\bf x}},
\end{equation}
where the coefficients $\{a_{\{n\}}\}$
of the $\{q\}$-th degree macroscopic structure polynomial
$y_{\{q\}}({\bf x};\tilde{\bf x})$
[$\equiv\sum_{\{n\}=\{0\}}^{\{q\}}a_{\{n\}}\prod_{i=1}^d (x_i-\tilde{x}_i)^{n_i}$
with the sum running over the set of indices $0\le n_i\le q_i$ for all $1\le i\le d$]
are determined by $\partial w^2_{\{q\}}/\partial a_{\{n\}}=0$.
Hence, the coefficients $\{a_{\{n\}}\}$ satisfy the following relation
\begin{equation}
\langle y({\bf x})\prod_{i=1}^d (x_i-\tilde{x}_i)^{n_i}
\rangle_{\prod_{i=1}^dl_i;\tilde{\bf x}}
=\sum_{\{n'\}=\{0\}}^{\{q\}}a_{\{n'\}}\left\{\prod_{i=1}^d\left(\frac{l_i}{2}
\right)^{n_i+n'_i} \frac{[1+(-1)^{n_i+n'_i}]}{2(n_i+n'_i+1)}\right\}.
\end{equation}
To determine $\{a_{\{n\}}\}$, one needs to deal with the inverse of a tensor
of rank $2d$.
By choosing a set of orthonormal functions as the basis,
we can actually avoid the difficulty in solving the inverse of a high-rank tensor.

Let $\{f_n(x);n=1,2,\ldots\}$ denote such a set of orthonormal functions on $[-1,1]$
(i.e., $\langle f_m(x)f_{m'}(x)\rangle_{[-1,1]}=\delta_{m,m'}$).
$y_{\{q\}}({\bf x};\tilde{\bf x})$ is then expressed as
\begin{equation}
y_{\{q\}}({\bf x};\tilde{\bf x})=\sum_{\{n\}=\{0\}}^{\{q\}}c_{\{n\}}\prod_{i=1}^d
f_{n_i} \left[\frac{2(x_i-\tilde{x}_i)}{l_i}\right].
\end{equation}
The requirement $\partial w^2_{\{q\}}/\partial c_{\{n\}}=0$ leads to
\begin{equation}
c_{\{n\}}=\left\langle y({\bf x})\prod_{i=1}^d f_{n_i}\left[
\frac{2(x_i-\tilde{x}_i)}{l_i}\right]\right\rangle_{\prod_{i=1}^dl_i;\tilde{\bf x}}.
\end{equation}
Therefore,
\begin{equation}
y_{\{q\}}({\bf x};\tilde{\bf x})=\sum_{\{n\}=\{0\}}^{\{q\}}\left\langle y({\bf x})
\prod_{i=1}^d f_{n_i}\left[\frac{2(x_i-\tilde{x}_i)}{l_i}\right]
\right\rangle_{\prod_{i=1}^dl_i;\tilde{\bf x}} \prod_{i=1}^d f_{n_i}
\left[\frac{2(x_i-\tilde{x}_i)}{l_i}\right].
\end{equation}
The above result can be viewed as the original $y({\bf x})$
convolved by some specific filter.
Based on the problems to be tackled, one can choose a specific set
of orthonormal functions to facilitate their analysis.
In this paper, we plan to analyze the scaling behaviors of the correlation function
and the detrended variances of the systems.
Thus, the set of Legendre polynomials\cite{Atlas87}
\begin{equation}
P_i(x)=\sum_{j=0}^{[i/2]}(-1)^j\frac{(2i-2j-1)!!}{(2j)!!(i-2j)!}x^{i-2j} ~~%
\mbox{for $i=1,2,\ldots$,}
\end{equation}
is a perfect candidate to be chosen.
From the orthogonal property
\begin{equation}
\langle P_i(x)P_{i'}(x)\rangle=\frac{\delta_{i,i'}}{2i+1},
\end{equation}
we then define our basis by
\begin{equation}
f_i(x)\equiv\sqrt{2i+1}P_i(x).
\end{equation}

Besides, in the following analysis we will use the symbol $[q]$
to denote the union of the sets of indices $\{n\}\equiv\{n_1,n_2,\ldots,n_d\}$
with $|\{n\}|\equiv n_1+n_2+\cdots+n_d\le q$ and all $n_i$ being nonnegative integers.
Hence,
\begin{equation}
y[q]({\bf x};\tilde{\bf x}) =\sum_{|\{n\}|\le q}\left\langle y({\bf x})
\prod_{i=1}^d P_{n_i}\left[\frac{2(x_i-\tilde{x}_i)}{l_i}\right]
\right\rangle_{\prod_{i=1}^dl_i;\tilde{\bf x}} \prod_{i=1}^d
\left\{(2n_i+1)P_{n_i}\left[\frac{2(x_i-\tilde{x}_i)}{l_i}\right]\right\}.
\end{equation}
As an illustration, let us consider the case with $d=2$.
With the dimensionless quantity $z_i\equiv(x_i-\tilde{x}_i)/l_i$,
we have the coefficients
\begin{eqnarray*}
c_{00}&=&\langle y({\bf x})\rangle_{l_1\times l_2;\tilde{\bf x}} \\
c_{10}&=&2\sqrt{3}\langle y({\bf x})z_1\rangle_{l_1\times l_2;\tilde{\bf x}} \\
c_{11}&=&12\langle y({\bf x})z_1z_2\rangle_{l_1\times l_2;\tilde{\bf x}} \\
c_{20}&=&\frac{\sqrt{5}}{2}\langle y({\bf x})(12z_1^2-1)
\rangle_{l_1\times l_2;\tilde{\bf x}} \\
c_{21}&=&\sqrt{15}\langle y({\bf x})(12z_1^2-1)z_2\rangle_{l_1\times l_2;\tilde{\bf x}}
\\
c_{30}&=&\sqrt{7}\langle y({\bf x})(20z_1^3-3z_1)\rangle_{l_1\times l_2;\tilde{\bf x}}
\end{eqnarray*}
and the corresponding $y[0]({\bf x};\tilde{\bf x})$ to $y[3]({\bf x};\tilde{\bf x})$
are also displayed as follows
\begin{eqnarray*}
y[0]({\bf x};\tilde{\bf x})&=&\langle y({\bf x})\rangle_{l_1\times l_2;\tilde{\bf x}} \\
y[1]({\bf x};\tilde{\bf x})&=&\langle y({\bf x})\rangle_{l_1\times l_2;\tilde{\bf x}}
+12\sum_{i=1}^2z_i\langle y({\bf x})z_i\rangle_{l_1\times l_2;\tilde{\bf x}} \\
y[2]({\bf x};\tilde{\bf x})&=&y[1]({\bf x};\tilde{\bf x})+\sum_{i=1}^2\left[\frac{5}{4}
(12z_i^2-1)\langle y({\bf x})(12z_i^2-1)\rangle_{l_1\times l_2;\tilde{\bf x}}\right]
+144z_1z_2\langle y({\bf x})z_1z_2\rangle_{l_1\times l_2;\tilde{{\bf x}}} \\
y[3]({\bf x};\tilde{\bf x})&=&y[2]({\bf x};\tilde{\bf x})
+\sum_{i=1}^2\left[7(20z_i^3-3z_i) \langle y({\bf x})(20z_i^3-3z_i)
\rangle_{l_1\times l_2;\tilde{\bf x}}\right. \\
&&\left.+15(12z_i^2-1)z_{i'}\langle y({\bf x})(12z_i^2-1)z_{i'}
\rangle_{l_1\times l_2;\tilde{\bf x}}\right]
\end{eqnarray*}
with $i'\equiv i(\mbox{mod 2})+1$.
Consequently, we have explicitly derived the exact expression
of the macroscopic structures of fluctuating systems in arbitrary dimensions.
Our result is generally applicable for any continuous systems
regardless of the shape of the macroscopic structures.
No more numerical fitting is needed.
Besides, the obtained result can greatly help further analytical investigation
of fluctuating systems.

\section{The Detrended Variances and the Correlation Function}

In the following, we would like to obtain the explicit and exact relations
between the detrended variance functions and the correlation function.
It will then facilitate the scaling analysis,
which will be undertaken in the next section. First,
the correlation function is defined as
\begin{equation}
G({\bf r})\equiv \overline{\left[y({\bf x})-y({\bf x}+{\bf r})\right]^2}
\end{equation}
with the overbar denoting the average over the whole system.
With some calculation, it is straightforward to obtain the relation
between the correlation function $G({\bf r})$ and the average
of the original variance function $\overline{w^2(\prod_{i=1}^dl_i;\tilde{\bf x})}$
(i.e., taking the average over all the windows of size $\prod_{i=1}^dl_i$):
\begin{equation}
\overline{w^2\left(\prod_{i=1}^dl_i;\tilde{\bf x}\right)}=\prod_{i=1}^d
\left[\frac{2}{l_i^2}\int_0^{l_i}dr_i(l_i-r_i)\right]\frac{1}{2}G({\bf r}).
\label{eq:WvsG}
\end{equation}
Next, the average of the $[q]$-th degree detrended variance function
is related to the average of the original variance function as:
\begin{eqnarray}
\overline{w^2[q]\left(\prod_{i=1}^dl_i;\tilde{\bf x}\right)} &=&
\overline{\langle\left[y({\bf x})-y[q]({\bf x};\tilde{\bf x})\right]^2
\rangle_{\prod_{i=1}^dl_i;\tilde{\bf x}}}  \nonumber \\
&=&\overline{\langle y^2({\bf x})\rangle_{\prod_{i=1}^dl_i;\tilde{\bf x}}}
-\sum_{|\{n\}|\le q}\overline{c^2_{\{n\}}}  \nonumber \\
&=&\overline{w^2\left(\prod_{i=1}^dl_i;\tilde{\bf x}\right)}
-\sum_{|\{n\}|=1}^q\overline{c^2_{\{n\}}}.  \label{eq:WQvsG}
\end{eqnarray}
Subsequently,
\begin{eqnarray}
&&\overline{c^2_{\{n\}\not=\{0\}}}  \nonumber \\
&=&\prod_{i=1}^d\left[\frac{(2n_i+1)}{l_i^2}\int_{-l_i/2}^{l_i/2}dx_i
\int_{-l_i/2}^{l_i/2}dx'_i P_{n_i}\left(\frac{2x_i}{l_i}\right)
P_{n_i}\left(\frac{2x'_i}{l_i}\right)\right] \overline{y(\tilde{\bf x}+{\bf x})
y(\tilde{\bf x}+{\bf x}')}  \nonumber\\
&=&-\frac{1}{2}\prod_{i=1}^d\left[\frac{(2n_i+1)}{l_i^2}\int_{-l_i/2}^{l_i/2}dx_i
\int_{-l_i/2-x_i}^{l_i/2-x_i}dr_i P_{n_i}\left(\frac{2x_i}{l_i}\right) P_{n_i}
\left(\frac{2x_i+2r_i}{l_i}\right)\right]G({\bf r}).
\end{eqnarray}
By employing the properties that the Legendre polynomials
are either even or odd functions of their arguments
and the correlation function $G({\bf r})$ is an even function of all $r_i$, we obtain
\begin{equation}
\overline{c^2_{\{n\}\not=\{0\}}}=-\frac{1}{2}\prod_{i=1}^d
\left\{\frac{2(2n_i+1)}{l_i^2}\int_0^{l_i}dr_i\left[ \int_{-l_i/2-r_i}^{l_i/2-r_i}dx_i
P_{n_i}\left(\frac{2x_i}{l_i}\right) P_{n_i}\left(\frac{2x_i+2r_i}{l_i}\right)\right]
\right\}G({\bf r}).  \label{eq:CNvsG}
\end{equation}
Substituting Eqs.\ (\ref{eq:WvsG}) and (\ref{eq:CNvsG}) into Eq.\ (\ref{eq:WQvsG}),
we consequently obtain the exact relation between the correlation function
and the average of the $[q]$-th degree detrended variance function as follows
\begin{equation}
\overline{w^2[q]\left(\prod_{i=1}^dl_i;\tilde{\bf x}\right)}=\frac{1}{2}\prod_{i=1}^d
\left(\frac{1}{l_i^2}\int_0^{l_i}dr_i\right)G({\bf r})K[q]({\bf r})\label{eq:WQvsKG}
\end{equation}
with the kernel given by
\begin{equation}
K[q]({\bf r})=\sum_{|\{n\}|\le q}\left\{\prod_{i=1}^d\left[2(2n_i+1)
\int_{-l_i/2}^{l_i/2-r_i}dx_iP_{n_i}\left(\frac{2x_i}{l_i}\right)
P_{n_i}\left(\frac{2x_i+2r_i}{l_i}\right)\right]\right\}.
\end{equation}

As an illustration, let us consider the case with $d=2$.
We explicitly list out $K[0]({\bf r})$ to $K[3]({\bf r})$:
\begin{eqnarray*}
K[0]({\bf r})&=&\prod_{i=1}^2(2l_i-2r_i), \\
K[1]({\bf r})&=&K[0]({\bf r})+\sum_{i=1}^2\left(2l_i-6r_i+4\frac{r_i^3}{l_i^2}\right)
(2l_{i'}-2r_{i'}), \\
K[2]({\bf r})&=&K[1]({\bf r})+\sum_{i=1}^2
\left(2l_i-10r_i+20\frac{r_i^3}{l_i^2}-12\frac{r_i^5}{l_i^4}\right)(2l_{i'}-2r_{i'})
+\prod_{i=1}^2\left(2l_i-6r_i+4\frac{r_i^3}{l_i^2}\right), \\
K[3]({\bf r})&=&K[2]({\bf r})+\sum_{i=1}^2\left[\left( 2l_i-14r_i+56\frac{r_i^3}{l_i^2}
-84\frac{r_i^5}{l_i^4}+40\frac{r_i^7}{l_i^6}\right)(2l_{i'}-2r_{i'})\right. \\
&&\left.+\left(2l_i-10r_i+20\frac{r_i^3}{l_i^2}-12\frac{r_i^5}{l_i^4}\right)
\left(2l_{i'}-6r_{i'}+4\frac{r_{i'}^3}{l_{i'}^2}\right)\right],
\end{eqnarray*}
with $i'\equiv i(\mbox{mod 2})+1$.
In a word, the correlation function and the $[q]$-th degree detrended variance function
are closely related through a kernel function.
Thus, before undertaking the scaling analysis of the fluctuating systems,
one should first investigate the properties of these kernel functions.

\section{Scaling Analysis}

In this section, we will first explore some important properties
of the kernel functions and then analyze the scaling relations among
$\overline{w^2(\prod_{i=1}^dl_i;\tilde{\bf x})}$,
$\overline{w^2[q](\prod_{i=1}^dl_i;\tilde{\bf x})}$, and $G({\bf r})$.
The kernel $K[q]({\bf r})$ under consideration can be recast as
$K[q]({\bf r})\equiv\sum_{|\{n\}|\le q}\left[\prod_{i=1}^dF_{n_i}(r_i)\right]$.
By some calculation, we first obtain the following property of $F_{n_i}(r_i)$:
with $m_i$ being any nonnegative number (including non-integers),
\begin{equation}
\int_0^{l_i}dr_ir_i^{2m_i}F_{n_i}(r_i)=\left\{
\begin{array}{ll}
0, & m_i=1,2,\ldots,n_i-1; \\
\frac{l_i^{2m_i+2}(-1)^{n_i}(2n_i+1)\Gamma(m_i+1)^2}
{(2m_i+1)\Gamma(m_i-n_i+1)\Gamma(m_i+n_i+2)}, & \mbox{otherwise.}
\end{array}
\right.
\end{equation}
We then will employ the above relation to compute
$\int_{\prod_{i=1}^d[0,l_i]}K[q]({\bf r})r^{2\beta}d{\bf r}$ (with $r\equiv|{\bf r}|$),
in order to explore the scaling relations
between the correlation function and the detrended variance functions.
For $\beta$ being a positive integer,
\begin{eqnarray*}
&&\int_{\prod_{i=1}^d[0,l_i]}K[q]({\bf r})r^{2\beta}d{\bf r} \\
&=&\sum_{|\{n\}|\le q}\sum_{|\{m\}|=\beta} \frac{\beta !}{\prod_{i=1}^d(m_i!)}
\left[\prod_{i=1}^d \int_0^{l_i}F_{n_i}(r_i)r_i^{2m_i}dr_i\right] \\
&=&\sum_{|\{n\}|\le q}\sum_{|\{m\}|=\beta} \frac{\beta !}{\prod_{i=1}^d(m_i!)}
\prod_{i=1}^d\left\{
\begin{array}{ll}
0, & m_i=1,2,\ldots,n_i-1; \\
\frac{l_i^{2m_i+2}(-1)^{n_i}(2n_i+1)\Gamma(m_i+1)^2}
{(2m_i+1)\Gamma(m_i-n_i+1)\Gamma(m_i+n_i+2)}, & \mbox{otherwise.}
\end{array}
\right.
\end{eqnarray*}
With some calculation and by induction, it can be shown that
the above integral is equal to 0 for $\beta=1,2,\ldots,q$.
In addition, in many physical situations, one usually takes the observation window
with equal side lengths (i.e., $l_i=l$ for all $i$) and, then,
the above integral is proportional to $l^{2\beta+2d}$, for $\beta=q+1,q+2,\ldots$.

If $\beta$ is not an integer, the multinomial expansion is no more applicable.
The calculation becomes extremely complicated
and no general closed form can be attained.
Since most physical situations are involved with the substrate dimension $d=2$,
we then choose to undertake the calculation for $d=2$
with the side lengths of the observation window $l_1=l_2=l$.
First, consider the integral
\begin{eqnarray}
&&H_\beta(l,n_1,n_2)  \nonumber \\
&\equiv&\int_{l\times l}d{\bf r}r_1^{n_1}r_2^{n_2}(r_1^2+r_2^2)^\beta \nonumber \\
&=&\frac{l^{n_1+n_2+2\beta+2}}{2(n_1+n_2+2\beta+2)}\sum_{i=1}^2
B\left(\frac{n_i+1}{2},-\beta-\frac{n_i+1}{2};\frac{1}{2}\right)\label{eq:Hbeta}
\end{eqnarray}
with the incomplete Beta function\cite{Atlas87}
\[
B(\nu,\mu;x)\equiv\int_0^xt^{\nu-1}(1-t)^{\mu-1}dt.
\]
By employing the above integral and the expansion formula for the Legendre polynomial,
we then derive
\begin{eqnarray}
&&\int_{l\times l}d{\bf r}F_{n_1}(r_1)F_{n_2}(r_2)(r_1^2+r_2^2)^\beta  \nonumber \\
&=&4l^2H_\beta(l,0,0)+4\prod_{i=1}^2 \left[\sum_{j_i=0}^{n_i}
\frac{(2n_i+1)(-n_i)_{j_i}(1+n_i)_{j_i}}{(2j_i+1)(j_i!)^2l^{2j_i}}\right]
H_\beta(l,2j_1+1,2j_2+1)  \nonumber \\
&&-4(2n_2+1)\sum_{k=0}^{n_2}\frac{(-n_2)_k(1+n_2)_k}{(2k+1)(k!)^2l^{2k-1}}
H_\beta(l,0,2k+1)  \nonumber \\
&&-4(2n_1+1)\sum_{j=0}^{n_1}\frac{(-n_1)_j(1+n_1)_j}{(2j+1)(j!)^2l^{2j-1}}
H_\beta(l,2j+1,0)  \label{eq:FvsH}
\end{eqnarray}
with $(n)_j\equiv n(n-1)\cdots(n-j+1)$.
Subsequently, by using Eqs.\ (\ref{eq:Hbeta}) and (\ref{eq:FvsH})
and the following relations for the incomplete Beta function\cite{Atlas87}
\begin{eqnarray}
B(\mu,\nu;x)&=&\frac{\Gamma(\mu)\Gamma(\nu)}{\Gamma(\mu+\nu)}-B(\nu,\mu;1-x)\\
B(\nu,\mu;x)&=&\frac{(1-x)^\mu}{\nu}\sum_{j=0}^\infty
\frac{(\mu+\nu)_j}{(\nu+1)_j}x^{j+\nu},
\end{eqnarray}
we finally obtain
\begin{eqnarray}
&&\int_{l\times l}K[q]({\bf r})(r_1^2+r_2^2)^\beta d{\bf r}  \nonumber\\
&=&\sum_{n_1+n_2\le q}\int_0^ldr_1\int_0^ldr_2F_{n_1}(r_1)F_{n_2}(r_2)
(r_1^2+r_2^2)^\beta  \nonumber\\
&=&l^{2\beta+4}\left\{\frac{2^\beta(q^2+2q+2)}{\beta+1} \sum_{\nu=0}^\infty
\frac{(-\beta)_\nu}{2^{\nu}(1/2)_{\nu+1}} +2^\beta\sum_{\nu=0}^\infty
\frac{(-\beta)_\nu}{2^\nu} \sum_{n_1+n_2\le q}(2n_1+1)(2n_2+1)\right.  \nonumber \\
&&\times\sum_{j=0}^\infty\sum_{k=0}^\infty
\frac{(-n_1)_j(1+n_1)_j(-n_2)_k(1+n_2)_k}{(2j+1)(2k+1)(j+k+\beta+2)(j!k!)^2}
\left[\frac{1}{(k+1)_{\nu+1}}+\frac{1}{(j+1)_{\nu+1}}\right]-2^{\beta+2}\nonumber\\
&&\left.\times\sum_{\nu=0}^\infty\frac{(-\beta)_\nu}{2^\nu}\sum_{n=0}^q(2n+1)(q-n+1)
\sum_{k=0}^\infty \frac{(-n)_k(1+n)_k}{(2k+1)(2k+2\beta+3)(k!)^2}
\left[\frac{1}{(k+1)_{\nu+1}}+\frac{1}{(1/2)_{\nu+1}}\right]\right\}\nonumber \\
&\propto&l^{2\beta+4}.
\end{eqnarray}

Next, we will proceed to analyze the scaling relations among the correlation function,
the average of the original variance function,
and the averages of the detrended variance functions.
For a fluctuating system consisting of a macroscopic structure,
we can recast $y({\bf x})$ as
$y({\bf x})=y_{\mathrm{macro}}({\bf x})+y_{\mathrm{sto}}({\bf x})$.
Here, $y_{\mathrm{macro}}({\bf x})$ represents the macroscopic structure,
which is expected to be continuous and smooth.
$y_{\mathrm{sto}}({\bf x})$ represents the stochasticity
relative to the macroscopic structure.
From the definition of the correlation function $G({\bf r})$,
$G({\bf r})$ can also be separated into two parts
$G_{\mathrm{macro}}({\bf r})+G_{\mathrm{sto}}({\bf r})$.
It is physically reasonable to assume that $G_{\mathrm{macro}}({\bf r})$ is analytic
and thus can be expressed in terms of a power series expansion.
Since the correlation function is a ``difference'' correlation function,
we further expect that $G_{\mathrm{macro}}({\bf r})$ is an even function
of all $r_i$ and thus
$G_{\mathrm{macro}}({\bf r})=\sum_{|\{n\}|=1}^\infty b_{\{n\}}\prod_{i=1}^dr_i^{2n_i}$.
From Eq.\ (\ref{eq:WvsG}), it is straightforward to see that
the average of the square of the original variance function will bear exactly
the same scaling behavior as the correlation function.
In most physical situations, the dominant term of $G_{\mathrm{sto}}({\bf r})$
is proportional to $r^{2\alpha}$ with a characteristic scaling exponent $\alpha$,
which is usually not an integer.
Thus, for the experimental systems with the formation of the macroscopic structure,
it is hard to measure the scaling exponent $\alpha$ directly,
if $r^{2\alpha}$ is not the leading term in the correlation function.
By substituting the above expansion of $G({\bf r})$ into Eq.\ (\ref{eq:WQvsKG})
and employing the properties of the kernel functions
(which we just derived in this section),
we rigorously show that, by raising the degree of the detrended variance function,
the contribution due to macroscopic structure can be successfully suppressed.
The above result is applicable in any dimensions.
In addition, at least in two dimensions, we rigorously show that
the detrended variance functions do retain the scaling exponent $\alpha$
(if it is not an integer) originated from the stochastic nature
of the fluctuating systems.

\section{Application in Interface Superroughening}

Dynamics of superrough growth processes can be modeled
by stochastic partial differential equations:
\begin{equation}
\frac{\partial h({\bf x},t)}{\partial t}=F(\{\nabla h\})+\eta({\bf x},t)+v_0
\end{equation}
with $h({\bf x},t)$ denoting the interfacial heights,
${\bf x}$ being a position vector in $d$ dimensional substrate,
$F$ describing the surface relaxation mechanism, $\eta$ representing the noise,
and $v_0$ denoting the average interface advancing velocity.
The global fluctuation of the interface can be measured by the global interfacial width
$w(L,t)$ (with $L$ denoting the linear size of the substrate), which is scaled as
\begin{equation}
w(L,t)\sim\left\{
\begin{array}{ll}
t^{\alpha/z}, & \mbox{for $t\ll L^z$;} \\
L^\alpha, & \mbox{for $t\gg L^z$.}
\end{array}
\right.
\end{equation}
Here, the two independent scaling exponents $\alpha$ and $z$ are given the names
the global roughness exponent and the dynamic exponent, respectively.
The superrough interface is defined by the global roughness exponent $\alpha>1$,
i.e., $w_{\mathrm{saturated}}(L)/L\to\infty$ as $L\to\infty$.
Qualitatively, the superrough growth processes gradually develop global mountains
or valleys as time increases.
By taking a snapshot of $(d+1)$-dimensional superrough interface morphology
at a certain time,
the formalism we just developed in the previous sections can be applied right away.
Besides, one of the most distinct features of superrough growth processes is that
the local scaling differs from the global scaling; namely, the local interfacial width
\begin{equation}
w(l,t)\sim\left\{
\begin{array}{ll}
t^{\alpha/z}, & \mbox{for $t\ll l^z$;} \\
t^{(\alpha-\alpha_{\mathrm{loc}})/z}l^{\alpha_{\mathrm{loc}}},
& \mbox{for $l^z\ll t\ll L^z$;} \\
L^{\alpha-\alpha_{\mathrm{loc}}}l^{\alpha_{\mathrm{loc}}}, & \mbox{for $t\gg L^z$}
\end{array}
\right.
\end{equation}
with $l$ denoting the linear size of the local observation window.
The exponent $\alpha_{\mathrm{loc}}$ is usually called the local roughness exponent.

In the following, we will discuss two important examples
in interfacial superroughening processes as applications.
The first example is the experiment of nickel surfaces grown
by pulse-current electrochemical deposition in 2+1 dimensions\cite{Saitou02}.
The surface morphology taken by atomic force microscopy clearly shows
the formation of large mounds.
The values of the scaling exponents, measured in experiments\cite{Saitou02},
are $\alpha=2.8$, $\alpha_{\mathrm{loc}}=1.0$, and $z=4.1$.
It is known that electrochemical noise can be regarded as
temporally correlated noise\cite{Roseler01}.
Thus, we propose the following stochastic partial differential equation
to describe this growth process:
\begin{equation}
\frac{\partial h({\bf x},t)}{\partial t}=-\nu\nabla^4h+\eta({\bf x},t)+v_0
\label{eq:ECgrowth}
\end{equation}
with the noise correlation
\begin{equation}
\langle\eta({\bf x},t)\eta({\bf x}',t')\rangle
=D|t-t'|^{2\theta-1}\delta({\bf x}-{\bf x}').
\end{equation}
In the RHS of Eq. (\ref{eq:ECgrowth}), the first term accounts for
the surface diffusion of deposits to the position of lowest chemical potential
and the second term accounts for the temporally correlated noise.
In 2+1 dimensions, by simple scaling analysis, we obtain $\alpha=1+4\theta$ and $z=4$.
In addition, generalizing the analysis in Ref. \cite{Pang04a} to 2+1 dimensions,
we obtain $\alpha_{\mathrm{loc}}=1$\cite{details}.
Thus, the pulse-current electrochemical deposition processes can be well described
by Eq.\ (\ref{eq:ECgrowth}) with the noise correlation index $\theta=0.45$.

Next, let us consider a broader class of growth processes in $d+1$ dimensions:
\begin{equation}
\frac{\partial h({\bf x},t)}{\partial t}=(-1)^{m+1}\nu\nabla^{2m}h+\eta({\bf x},t)+v_0
\end{equation}
with $m=1,2,3,\ldots$.
The noise can be either the white noise
$\langle\eta({\bf x},t)\eta({\bf x}',t')\rangle=D\delta(t-t')\delta({\bf x}-{\bf x}')$
or the correlated noise $\langle\eta({\bf x},t)\eta({\bf x}',t')\rangle
=D |t-t'|^{2\theta-1}|{\bf x}-{\bf x}'|^{2\rho-d}$.
The noise correlation indices $\rho$ and $\theta$ are restricted in the range $(0,1/2)$
to avoid the unphysical divergence.
Generalizing the analysis in Refs. \cite{Pang04a} and \cite{Pang01} to $d+1$ dimensions,
we obtain $(\alpha,z)=(\frac{2m-d}{2},2m)$ for the cases with white noise
and $(\alpha,z)=(\frac{2m-d}{2}+\rho+2m\theta,2m)$ for the cases with correlated noise.
Moreover, for the above class of growth processes with $\alpha>1$,
the local roughness exponent $\alpha_{\mathrm{loc}}$ is always equal to 1,
independent of $d$, $m$, $\rho$, and $\theta$\cite{details}.
Is this just a coincidence?
The further analytical calculation shows that for the above mentioned growth processes
with $\alpha>1$, $G_{\mathrm{macro}}({\bf r},t)$ (the part of the correlation function
attributed to the macroscopic structure formation) takes the form
$\sum_{q=1}^\infty A_{2q}t^{2(\alpha-q)/z}r^{2q}$ and
$G_{\mathrm{sto}}({\bf r},t)\sim r^{2\alpha}$ for $r^z\ll t\ll L^z$\cite{details}.
This result confirms our scaling arguments in the previous section.
Note that the leading term of $G({\bf r},t)$ is $t^{2(\alpha-1)/z}r^2$
for $r^z\ll t\ll L^z$.
Recall that the square of the the local interfacial width bears
the same scaling behaviors as the correlation function.
Consequently, we have $\alpha_{\mathrm{loc}}=1$,
an intriguing property of superrough interfaces.
To successfully extract the scaling exponent $\alpha$
from the experimentally measured data of interface morphology,
the detrended variance function of degree $[q]$
with the integer $q>\alpha-1$ needs to be employed.

\section{Conclusion}

In conclusion, fluctuating systems with the formation of macroscopic structures
are commonly observed in Nature, for example, growth by molecular beam epitaxy,
growth by electrochemical deposition, cultivated brain tumor growth,
spontaneous imbibition of viscous fluids in porous media,
the mosaic structure of DNA sequences, physiology signals,
and the clustering features of seismic sequences.
The ``generalized detrended fluctuation analysis'' method is designed
to determine the correct scaling behaviors of fluctuations
in the presence of possible trends (or macroscopic structures)
without knowing their origin and shape.
In this paper, we undertake an extensive study on this method in arbitrary dimensions.
We not only derive the explicit expression of the macroscopic structures,
but also obtain the exact relations between the correlation function
and the detrended variance functions.
Through the rigorous analysis of the kernel functions, we explicitly show that
the detrended variance functions can successfully suppress the influence
of the macroscopic structures on the scaling behaviors
and, at the same time, keep the scaling behaviors of fluctuations unaltered.
Finally, as applications, we discuss some important examples
in interfacial superroughening and give explanations
about an intriguing property of superrough interfaces $\alpha_{\mathrm{loc}}=1$.
All the obtained results are exact and applicable for continuous fluctuating systems
in arbitrary dimensions.

\begin{acknowledgments}
The work of N.-N. Pang is supported in part by the National Science Council
of the Republic of China under Grant No. NSC-94-2112-M002-021.
The work of W.-J. Tzeng is supported in part by the National Science Council
of the Republic of China under Grant No. NSC-94-2112-M032-008.
\end{acknowledgments}


\begin{thebibliography}{99}
\bibitem{details} The detailed mathematical derivation will be published elsewhere.
\end{thebibliography}
\end{document}